\documentclass[letterpaper,11pt,twocolumn]{article}

\usepackage{amsmath,amsfonts,amssymb}      %
\usepackage{mathtools}                     %
\usepackage{bm}                            %
\usepackage{slashed}                       %

\usepackage[main=english]{babel}           %
\usepackage{microtype}                     %
\usepackage[strict]{changepage}            %
\usepackage{url}                           %
\usepackage{titling}                       %
\usepackage{fancyhdr}                      %
\usepackage{titlesec}                      %
\usepackage[utf8]{inputenc}
\usepackage{lettrine}                      %
\usepackage{lastpage}
\usepackage{tocloft}
\usepackage[noblocks,affil-it]{authblk}
\usepackage[svgnames,x11names]{xcolor}     %
\usepackage{tikz}                          %
\usepackage{tikzrput}                      %

\usepackage{enumitem}                      %
\usepackage{etaremune}                     %

\usepackage{dcolumn}                       %
\usepackage{multirow}                      %

\usepackage{graphicx}                      %
\usepackage[font=small]{caption}           %

\usepackage{calc}                          %
\usepackage[margin=20mm]{geometry}
\usepackage{newtxtext,newtxmath}

\usepackage[colorlinks=true,linkcolor=MediumBlue,citecolor=MediumBlue,urlcolor=MediumBlue]{hyperref}
\newcommand{\HorRule}{\color{DarkGoldenrod}\rule{\linewidth}{1pt}}

\pretitle{
  \vspace{-50pt}
  \HorRule\vspace{0pt}
  \fontsize{28}{34}\selectfont\color{Black}
}

\posttitle{\par\vskip 15pt}

\preauthor{} 

\postauthor{
  \vspace{0pt}
  \par\HorRule
  \vspace{-30pt}
}

\fancypagestyle{first}{\fancyhf{}\fancyfoot[R]{}}
\fancypagestyle{foot} {\fancyhf{}\fancyfoot[R]{\footnotesize Page \thepage\ of \pageref{LastPage}}}

\graphicspath{{.}{figures/}}

\titlespacing{\section}{0pt}{14pt}{5pt}

\makeatletter
\renewcommand{\l@section}{\@dottedtocline{0}{0pt}{15pt}}
\makeatother

\renewcommand{\thesection}{\arabic{section}.}

\usepackage{csquotes}
\usepackage[style=texfiles/bibstyle,
            biblabel=brackets,
            chaptertitle=false,
            articletitle=true,
            pageranges=false,
            eprint=false,
            backend=biber,
            sorting=none,
            maxnames=5,           %
            minnames=5            %
            ]{biblatex}

\DefineBibliographyStrings{english}{andothers = {\textit{et al}\adddot}}

\DeclareFieldFormat[article]{title}{\mkbibemph{#1}} 

\usepackage{pdfpages}

\begin{document}

\title{Opportunities for Nuclear Physics \& Quantum Information Science}

\preauthor{
\rput[tr](17.58,5.3){\includegraphics[width=0.5\columnwidth,clip=true]{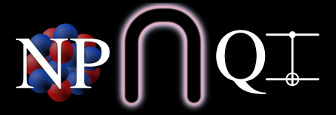}}
\textbf{Editors}\\
Ian C. Clo\"et$^1$ and Matthew R. Dietrich$^1$ \\[1.0em]
\textbf{Contributors} \\
}

\author[ 1]{Ian C. Clo\"et}
\author[ 1]{Matthew R. Dietrich}
\author[ 1]{John Arrington}
\author[2,3]{Alexei Bazavov}
\author[ 1]{Michael Bishof}
\author[ 1]{Adam Freese}
\author[4,5]{Alexey V. Gorshkov}
\author[6,7]{Anna Grassellino}
\author[ 1]{Kawtar Hafidi}
\author[ 8]{Zubin Jacob}
\author[ 9]{Michael McGuigan}
\author[10]{Yannick Meurice}
\author[ 1]{Zein-Eddine Meziani}
\author[ 1]{Peter Mueller}
\author[11]{Christine Muschik}
\author[12]{James Osborn}
\author[13]{Matthew Otten}
\author[14]{Peter Petreczky}
\author[15]{Tomas Polakovic}
\author[16]{Alan Poon}
\author[17]{Raphael Pooser}
\author[18]{Alessandro Roggero}
\author[19]{Mark Saffman}
\author[20]{Brent VanDevender}
\author[ 4]{Jiehang Zhang}
\author[21]{Erez Zohar}

\affil[ 1]{Physics Division, Argonne National Laboratory, Argonne, IL 60439 USA}
\affil[ 2]{Department of Computational Mathematics, Science and Engineering, Michigan State University, East Lansing, MI 48824, USA}
\affil[ 3]{Department of Physics and Astronomy, Michigan State University, East Lansing, MI 48824, USA}
\affil[ 4]{Joint Quantum Institute, NIST/University of Maryland, College Park, MD 20742, USA}
\affil[ 5]{Joint Center for Quantum Information and Computer Science, NIST/University of Maryland, College Park, MD 20742, USA}
\affil[ 6]{Fermi National Accelerator Laboratory, Batavia IL, 60510, USA}
\affil[ 7]{Northwestern University, Evanston, IL, 60208, USA}
\affil[ 8]{Birck Nanotechnology Center and Purdue Quantum Center, School of Electrical and Computer Engineering, Purdue University, West Lafayette, IN 47906, USA}
\affil[ 9]{Computational Science Initiative, Brookhaven National Laboratory, Upton, NY 11973, USA}
\affil[10]{Department of Physics and Astronomy, The University of Iowa, Iowa City, IA 52242, USA}
\affil[11]{Institute for Quantum Computing and Department of Physics and Astronomy, University of Waterloo, West Waterloo, Ontario, Canada}
\affil[13]{ALCF, Argonne National Laboratory, Argonne, IL 60439 USA}
\affil[13]{Nanoscience and Technology Division, Argonne National Laboratory, Argonne, IL 60439 USA}
\affil[14]{Physics Department, Brookhaven National Laboratory, Upton, NY 11973, USA}
\affil[15]{Physics and Material Science Divisions, Argonne National Laboratory, Argonne, IL 60439 USA}
\affil[16]{Institute for Nuclear and Particle Astrophysics and Nuclear Science Division, Lawrence Berkeley National Laboratory, Berkeley, CA 94720, USA}
\affil[17]{Quantum Information Science Group, Computational Sciences and Engineering Division, Oak Ridge National Laboratory, Oak Ridge, TN 37831, USA}
\affil[18]{Institute for Nuclear Theory, University of Washington, Seattle, WA 98195, USA}
\affil[19]{Department of Physics, University of Wisconsin, Madison, WI 53706, USA}
\affil[20]{Pacific Northwest National Laboratory, Richland, WA 99354, USA}
\affil[21]{Max Planck Institute of Quantum Optics, Hans-Kopfermann-Straß e 1, 85748 Garching, Germany}
\date{}

\begin{titlepage}
\includepdf{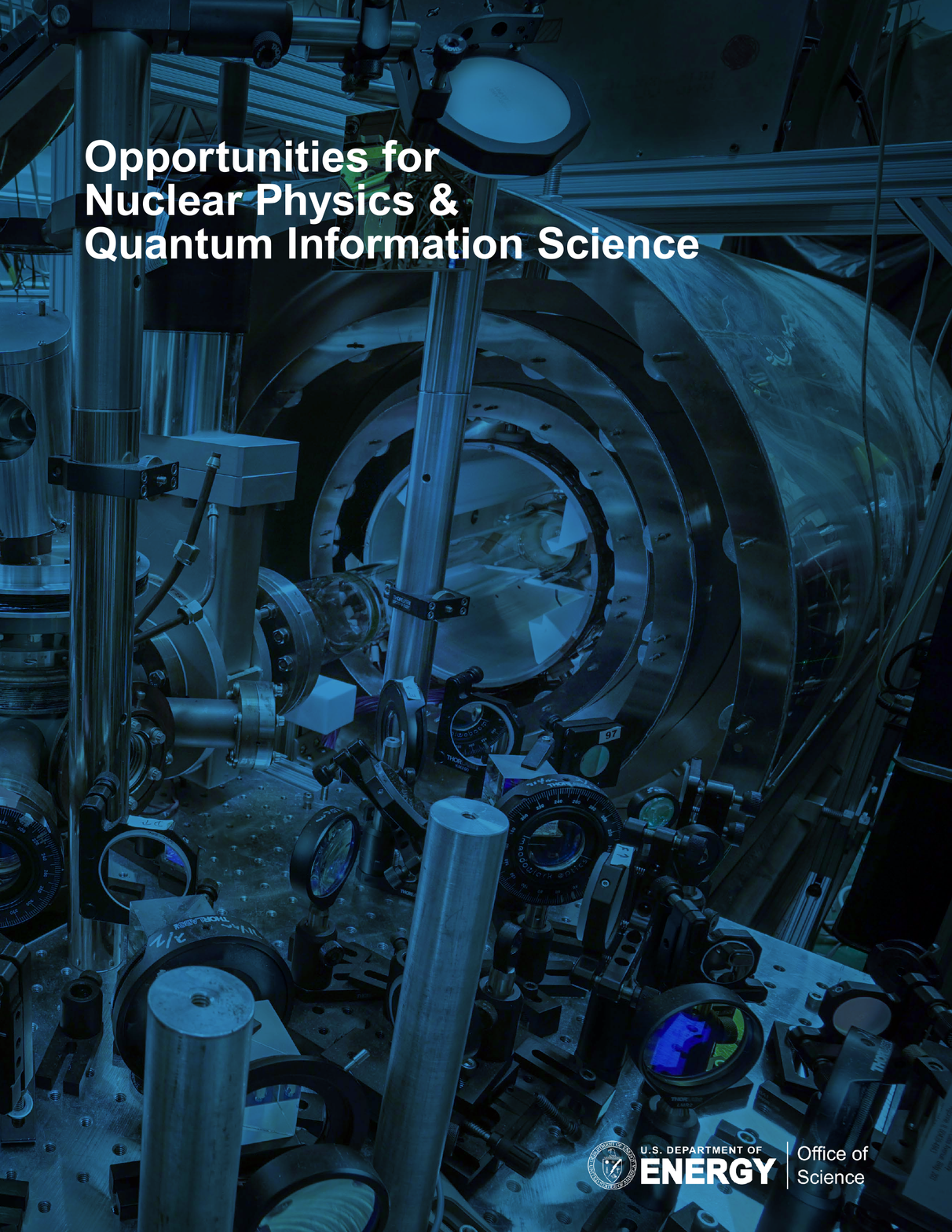}
\includepdf{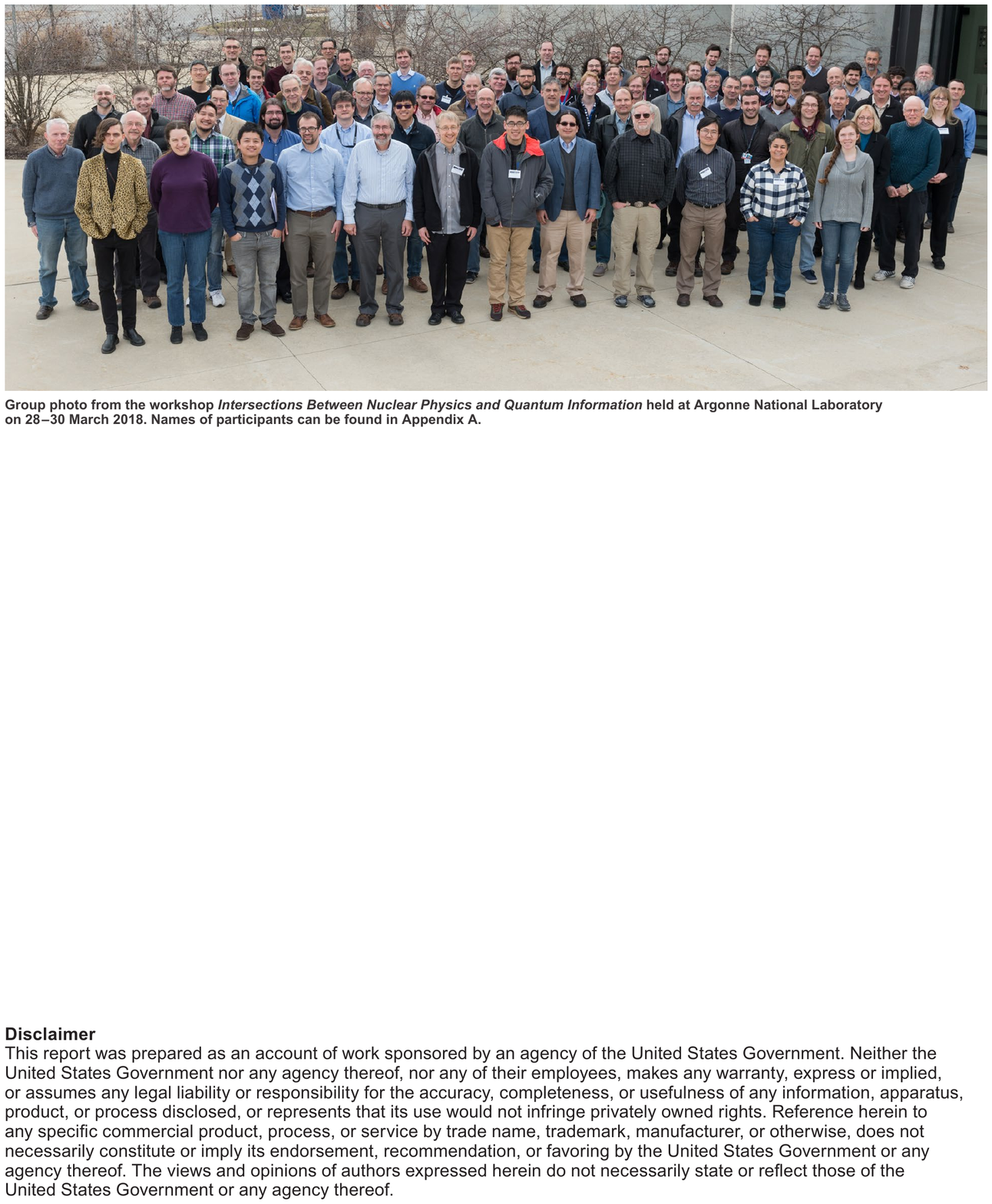}
\end{titlepage}
\setcounter{page}{3}

\twocolumn[
\begin{@twocolumnfalse}
  \maketitle
  \vspace*{-20pt}
  \begin{adjustwidth}{30pt}{30pt}\small
  \lettrine{T}{~~}his whitepaper is an outcome of the workshop {\it Intersections between Nuclear Physics and Quantum Information} held at Argonne National Laboratory on 28--30 March 2018 
\text{[}{\url{www.phy.anl.gov/npqi2018/}}{}\text{]}. The workshop brought together 116 national and international experts in nuclear physics and quantum information science to explore opportunities for the two fields to collaborate on topics of interest to the U.S. Department of Energy (DOE) Office of Science, Office of Nuclear Physics, and more broadly to U.S. society and industry. The workshop consisted of 22 invited and 10 contributed talks, as well as three panel discussion sessions. Topics discussed included quantum computation, quantum simulation, quantum sensing, nuclear physics detectors, nuclear many-body problem, entanglement at collider energies, and lattice gauge theories.
  \end{adjustwidth}
  \vspace*{15pt}
  \thispagestyle{first}
\end{@twocolumnfalse}]

\renewcommand\contentsname{}
\setcounter{tocdepth}{2}
\tableofcontents
\thispagestyle{foot}

\vspace{1.5em}
\addcontentsline{toc}{section}{Executive Summary}
\noindent\textbf{\Large Executive Summary}\\[0.3em]
Quantum information science (QIS) is poised to have a transformational impact on science and technology. Given the significant remaining technical and theoretical challenges the realization of this potential will require the cooperative efforts of academia, national laboratories, and industry so that a full spectrum of quantum technologies can be made ready for production use on real-world problems.  

Such a program can support the core mission of the Office of Nuclear Physics (ONP) by, e.g., the simulation of gauge field theories using quantum computers and enabling new nuclear physics experiments using quantum sensors. There is significant overlap in the theoretical and technical expertise between the NP and QIS communities, and many currently intractable problems in NP can potentially be tackled using technologies from QIS (e.g. quantum computing). By developing strategies that lead to better communication and collaboration between these two communities, ONP can play a leadership role in developing quantum technologies and advancing QIS theory. This whitepaper identifies five broad research opportunities that will help facilitate QIS research in the context of NP.

\textbf{Research Opportunity I:} \textit{To best leverage the respective strengths of universities, national laboratories, and industry, funding opportunities that encourage collaborative projects between the NP and QIS communities should be emphasized.}
The field of QIS is highly competitive, with many well-established research groups. Therefore, NP faces a challenge to invest strategically so that it can most effectively contribute to and benefit from QIS. Some relevant NP strengths include: superconducting technologies, microfabrication, supercomputing expertise, engineering expertise, isotope programs, and nuclear theory. Collaborations within national laboratories should also be encouraged, since the interdisciplinary nature of these institutions has a lot to offer QIS, which itself is highly interdisciplinary.  

\textbf{Research Opportunity II:} \textit{A broad theory program should be supported, which can, e.g., develop methods to address problems in NP using digital quantum computers and quantum  simulators, utilize QIS concepts to better understand nuclear phenomena (such as the nuclear many-body problem and hadronization), and develop new QIS applications of importance to nuclear physics.}
Technological development is often driven by the need to address important problems. Many facets of the nuclear many-body problem and quantum chromodynamics can not (currently) be addressed with conventional computing methods -- despite decades of effort. Strong interaction nuclear physics therefore provides an ideal application for QIS. However, significant advances in both NP and QIS theory (e.g. quantum algorithms) are still needed, which will require the dedicated effort of theorists from both communities. 

\textbf{Research Opportunity III:} {\it Support should be given to research that seeks to develop or capitalize on QIS technology with nuclear physics applications.} Development of technology originally intended for QIS has frequently generated detectors that can be used to advance basic research. Support for research intended to implement quantum sensors for nuclear physics applications would further the interests of both the NP and QIS communities, while helping to generate crucial collaborations.

\textbf{Research Opportunity IV:} \textit{Support should be given to develop a QIS workforce in the context of NP by funding graduate students and postdocs that conduct research at the intersection of these fields.}
NP problems that seem intractable today can potentially be addressed by a new NP workforce skilled in quantum computing and other aspects of QIS. To develop a robust QIS program within NP, a workforce skilled in both fields is therefore critical. Funding graduate students and postdocs should be a priority, where collaboration with the DOE Office of Advanced Scientific Computing Research (ASCR) should be encouraged, and support for an industry internship program may be considered to better respond to the demands of industry.

\textbf{Research Opportunity V:} \textit{Meetings such as workshops, schools, and seminars should be supported so that the NP and QIS communities can better understand the challenges and capabilities of their counterparts.}
The full extent of cross-pollination between NP and QIS is not yet fully understood. It is crucial to address this by improving general comprehension between these fields. A combination of workshops, pedagogical schools, and high-level seminars (all of which should be broadcast online) would promote the formation of new relationships and collaborations between interested parties in the NP and QIS communities. In addition a NP-QIS alliance could be formed to help disseminate pertinent information (e.g. job openings, seminars, workshops, etc) and thereby help form a more cohesive NP-QIS community. These are important elements for the success of a QIS program within ONP.

\section{Introduction}
Quantum information science (QIS) encompasses fields such as quantum computing and simulation, quantum sensors, quantum information theory, and quantum communication. By studying and harnessing the quantum nature of matter QIS has the potential for a transformational impact on science and technology.\footnote{This situation is not unlike the development of the transistor in the first half of the 20th century, which was also an outcome of quantum mechanics, and led to {\it inter alia} modern computing.} In the context of the U.S Department of Energy's (DOEs) Office of Nuclear Physics (ONP) program, which has the goal to ``{\it discover, explore, and understand all forms of nuclear matter}''\cite{ONP:mission}, the significant intersections between QIS and NP are evident, e.g., in computation, detectors, and quantum theory.

NP presents science with numerous important challenges, foremost amongst these is understanding the full implications of quantum chromodynamics (QCD), which is coupled to problems such as the nuclear many-body problem, stellar nucleosynthesis, and neutron stars. QCD is the asymptotically free gauge field theory that forms the strong interaction piece of the Standard Model of particle physics, and therefore provides the basis for understanding of all forms of nuclear matter. While QCD is easy to formulate in terms of quark and gluon fields~\cite{Altarelli:1981ax}, it is notoriously difficult to solve~\cite{Wilson:1974sk}. However, QIS provides new methods with which to study and understand aspects of NP, e.g., in principle quantum computing and quantum simulation provide the means to study aspects of QCD that are inaccessible with current computational techniques~\cite{Wiese:2013uua,Zohar:2015hwa}; quantum sensors may enable new and more precise measurements of nuclear phenomena~\cite{Degen:2017aa}; and quantum information theory provides a new lens with which to view and study entangled nuclear systems~\cite{Pichler:2015yqa} such as nuclei and the quark-gluon structure of QCD bound states and reactions.

Solving QCD will have profound implications for our understanding of the natural world, most notably it will explain how 99\% of the mass in the visible universe is created~\cite{NAP25171}. The infrared domain of QCD -- where the running coupling becomes large -- is also characterized by several interesting emergent phenomena, such as, quark-gluon confinement~\cite{Wilson:1974sk,Jaffe:1975fd}, dynamical chiral symmetry breaking~\cite{Pagels:1978ba,Weinberg:1978kz}, and hadron formation~\cite{Fodor:2012gf,Metz:2016swz}, which cannot be addressed using the perturbative methods that have proven so successful with other fundamental quantum field theories such as quantum electrodynamics.

The only known method to solve QCD over all energy-scales is lattice QCD, where QCD path integrals are evaluated on a 4-dimensional Euclidean space-time lattice by brute-force numerics. With the continual increase in conventional computing power lattice QCD is now having a dramatic impact on strong interaction physics. For example, calculations are now done at physical quark masses with large lattice volumes~\cite{Chang:2018uxx}, gluonic components of matter are being investigated~\cite{Alexandrou:2017oeh}, and the quark-gluon structure of light-nuclei is beginning to be explored~\cite{Winter:2017bfs}. With the arrival of the exascale era in conventional computing over the next few years~\cite{Habib:2016sce} lattice QCD will continue to increase its impact on QCD and hadron physics. However, there remain key problems in QCD that cannot be addressed with any known {\it ab initio} methods, foremost amongst these are arguably QCD at finite density~\cite{deForcrand:2010ys} and quark-gluon hadronization~\cite{Panero:2014qxa}. Quantum computing can potentially address these phenomena by overcoming signal-to-noise problems and providing a means to study real-time dynamics in quantum field theories~\cite{Lloyd:1996aa,Wiese:2014rla}.

\subsection{Noisy Intermediate-Scale Quantum Computing}
Quantum computing has caught the imagination of many since Richard Feynman famously stated ``{\it Nature is quantum \ldots So if we want to simulate it, we need a quantum computer}''~\cite{Feynman:1981aa,Feynman:1981tf}. The first theoretical proof that quantum computing is a possibility came a year or so earlier when Paul Benioff, from Argonne National Laboratory, developed the first quantum mechanical (Hamiltonian) model for a Turing machine, and established that a quantum Turing machine can be used to simulate a classical computer~\cite{Benioff:1981aa,Benioff:1982aa,Benioff:1982bb}. This problem was also recognized by Manin around the same time~\cite{Manin:1980aa}.

Quantum computers directly exploit quantum mechanical concepts such as superposition and entanglement, and therefore operate in a fundamentally different manner than classical computers. This allows quantum algorithms to be developed, which opens completely new possibilities in computer science by using processes fundamentally unavailable to classical computers. Numerous quantum algorithms already exist, where many have been proven to solve certain classes of problems exponentially faster than existing algorithms for conventional computers~\cite{Shor:1994jg}, and a wider range of problems with polynomial speed increases~\cite{Grover:1996rk}. A fundamental reason for this exponential increase in computational power is because of the quantum entanglement between quantum bits (qubits), which allows an exponential increase in information density~\cite{Rotta:2017aa}. The increase in computation power promised by quantum computers, and the ability to develop completely new types of quantum algorithms, opens the possibility that QIS can address numerous problems in nuclear physics that are currently computationally challenging or even intractable. 

Industry is now exploiting QIS to develop quantum computers and other technologies such as quantum sensors. Technology companies like Google~\cite{Google:QC}, IBM~\cite{IBM:QC}, Intel~\cite{Intel:QC}, and Microsoft~\cite{Microsoft:QC} have unveiled quantum computers with order 50 qubits (without error correction). In 2016 IBM began offering cloud access to its prototype quantum computers, and now provides access to two 5-qubit, one 16-qubit, and two 20-qubit processors through the IBM Q Experience~\cite{IBM:QC}. Rigetti~\cite{Rigetti:QC} also offers cloud based quantum computing services, and recently announced a 128-qubit processor with plans to make it available by August 2019~\cite{Rigetti:QC2}. Companies such as IonQ~\cite{IonQ:QC} are building quantum computers based on trapped ions, which currently offer much lower error rates than their superconducting counterparts, and D-wave~\cite{Dwave:QC} has built a quantum annealer consisting of 2000 qubits which is suited to tackling certain types of complex optimization problems. 

Quantum computing is now a reality, with machines of 50--100 (non-error corrected) qubits expected to become standard within the next few years, although these machines will still be limited by low circuit depth. This represents a significant technical milestone, although quantum computers of this size are still a long way from what is needed for many applications, such as to break RSA encryption using Shor's algorithm~\cite{Shor:1994jg}.\footnote{A realistic implementation of Shor's algorithm -- that includes error correction -- would currently require millions of physical qubits~\cite{OGorman:2017aa}.} The current era in quantum computing has therefore been labeled {\it Noisy Intermediate-Scale Quantum} (NISQ)~\cite{Preskill:2018aa} because of the relatively small number of qubits and the lack of error correction.

While quantum computers in the NISQ era may be well short of what is needed to reach {\it quantum supremacy}~\cite{Terhal:2018aa,Yung:2018aa} over a broad range of applications, a quantum computer with around 50--100 qubits cannot be simulated with contemporary supercomputers~\cite{Pednault:2017aa}. This suggests that quantum computers in the NISQ era may be approaching the sophistication needed to perform computational tasks that would be impossible on any classical computer for specific applications. Contemporary quantum computers are already finding important applications in fields such as quantum chemistry, and it remains an important challenge to determine what problems NISQ computers can have a strong impact on for nuclear physics.

\subsection{Nuclear Many-Body Problem}
The nuclear many-body problem~\cite{Ring:1980aa} is a key challenge for NP theory. Numerical solutions to this problem with current approaches, such as Green function and variational Monte Carlo methods~\cite{Carlson:2014vla}, are hindered by an exponential growth in quantum states with increasing number of nucleons~\cite{Pieper:2001mp}, together with a fermion sign problem~\cite{Troyer:2004ge,Li:2018wnz}. However, with quantum computers the computational cost of simulating quantum many-body systems can potentially be reduced from exponential to polynomial~\cite{Nielsen:2010aa}. Such approaches are already finding application in quantum chemistry~\cite{Lanyon:2010aa,Colless:2018aa}. 

The development of new quantum algorithms that have application for the nuclear many-body problem is already underway within NP. For example, Ref.~\cite{Kaplan:2017ccd} presents a spectral combing algorithm for finding the ground-state wave function of a quantum many-body system that does not require an initial trial wave function with good ground-state overlap; a quantum algorithm to calculate the dynamic linear response function of a quantum system is presented in Ref.~\cite{Roggero:2018hrn}; and a hybrid quantum-classical algorithm for studying the time evolution of out-of-equilibrium thermal states is given in Ref.~\cite{Lamm:2018siq}.

A first step toward a realistic nuclear physics calculation was taken in Ref.~\cite{Dumitrescu:2018njn}, which used the IBM QX5 and Rigetti 19Q cloud quantum computing resources to determine the binding energy of the deuteron to within a few percent. A tailored leading-order pionless effective field theory Hamiltonian provided the theoretical framework for the calculation. The deuteron ground-state was obtained using the variational quantum eigensolver algorithm (VQE)~\cite{Peruzzo:2014aa} with a low-depth version of the unitary coupled-cluster ansatz (UCC)~\cite{McClean:2016aa}. The VQE algorithm is a hybrid method where a quantum computer determines the energy expectation value for a wave function ansatz, then a classical optimizer supplies new parameter guesses for the wave function ansatz, and that continues until convergence. The UCC ansatz for the wave function has been shown to be exponentially faster on a quantum computer than on a classical computer~\cite{Peruzzo:2014aa}. Using these techniques and only a few qubits on a quantum computer the deuteron's ground-state binding energy was found to be within a few percent of the empirical value.

\subsection{Field Theories and Quantum Computing}
The best currently available method to calculate static properties of hadrons -- starting from the QCD Lagrangian -- is Wilson's lattice QCD. However, such a formulation does not provide access to real-time dynamics and systems at finite density~\cite{Wiese:2014rla,Silvi:2016cas}. Quantum computing can potentially overcome this restriction and significant progress has already been made in this direction. For example, simple bosonic~\cite{Retzker:2005aa,Jordan:2011ne} and fermionic~\cite{Cirac:2010aa,Mazza:2012aa} field theories have been formulated such that they could be studied using a quantum computer, and quantum link models provide an alternative non-perturbative formulation of non-abelian gauge theories that may allow QCD to be studied on a quantum computer~\cite{Wiese:2013uua,Zohar:2015hwa}. 

An important first step in the simulation of lattice gauge theories using a quantum computer was reported in Ref.~\cite{Martinez:2016yna}. Using a few-qubit analog quantum simulator\footnote{See Sect.~\ref{sec:qc} for a discussion of the different types of quantum computers.} the real-time dynamics of spontaneous particle-antiparticle creation in the vacuum was studied in 1+1 quantum electrodynamics (QED), the so called Schwinger model~\cite{Schwinger:1962tp,Kogut:1974ag}. An analog quantum simulator -- as opposed to a universal quantum computer -- is specifically designed to simulate a certain class of physical theories by, e.g., tuning the potential between atoms or ions in an optical lattice to mimic the physics of a particular Hamiltonian. The real-time evolution of the theory can then be studied by measuring the state of the analog quantum simulator. The Schwinger model has also recently been studied using IBM's cloud quantum computing resources in Ref.~\cite{Klco:2018zqz}, using a hybrid quantum-classical algorithm where classical computation is used to find symmetry sectors on which the quantum computer evaluates the dynamics of quantum fluctuations. The ground state of this model has also been studied with a variational
quantum simulator~\cite{Kokail:2018eiw}.

Analog quantum simulators with up to 50 trapped ions or atoms (qubits) have already been built~\cite{Zhang:2017aa,Bernien:2017aa}, and with their low error rates there exists significant opportunities for the further study of lattice gauge theories using these devices~\cite{Wiese:2013uua,Zohar:2015hwa}. In particular, with relevant expertise in atom trapping the development of an analog quantum simulator program to study lattice field theories within NP would be a real possibility.

\subsection{Entanglement at collider energies}
Concepts developed or extended within the field of quantum information theory, such as entanglement and quantum entropy, are beginning to impact the interpretation and understanding of the quark-gluon structure of QCD bound states and reactions. For example, Ref.~\cite{Kharzeev:2017qzs} considers the von Neumann entropy of the system of
partons (in this case gluons) resolved by deep inelastic scattering at small Bjorken $x$. Their analysis motivates the interpretation that this von Neumann entropy can be interpreted as the entropy of entanglement between the spatial region probed by deep inelastic scattering and the rest of the target. If this interpretation is correct then the entanglement entropy of a proton or nucleus could be studied using deep inelastic scattering, where event-by-event measurement of the hadronic final state is made. Such an experiment would be possible at the proposed electron-ion collider~\cite{NAP25171}.

Another interesting application is the impact of entanglement on thermalization in $e^+e^-$~\cite{Berges:2017zws,Berges:2017hne} and heavy ion~\cite{Baker:2017wtt,Bellwied:2018gck} collisons. In both these reactions it is observed that the final state particle spectra have thermal properties, even though re-scatterings effects in the final state appear not to be frequent enough to explain this. An interesting alternative explanation is that this thermalization is because of quantum entanglement between the final state particles. Interestingly, such effects have already been observed in recent measurements of isolated quantum many-body systems made of ultracold atoms~\cite{Kaufman:2016aa}.

Ideas related to quantum entanglement are also impacting the interpretation of hadron tomography~\cite{Hagiwara:2017uaz,Mulders:2018mmg,Mulders:2018ion}, parton-parton scattering~\cite{Liu:2018gae}, and condensed matter systems~\cite{Cervera-Lierta:2017tdt,Laflorencie:2015eck}. The interaction of quantum information theory and nuclear theory therefore provides interesting new opportunities to understand and interpret strong interaction phenomena.

\subsection{Document Outline}
This introduction attempts to provide a snapshot of QIS in the context of NP. A standard texbook on QIS that provides background to some of the terms and concepts discussed in this document is Ref.~\cite{Nielsen:2010aa}. This white paper on ``Opportunities for Nuclear Physics \& Quantum Information Science'' has been prepared at the request of the DOE Office of Science, Office of Nuclear Physics. It represents the outcome of the workshop {\it Intersections between Nuclear Physics and Quantum information} held at Argonne National Laboratory on 28--30 March 2018 [\url{www.phy.anl.gov/npqi2018/}]. This document is a complement to the white paper ``Quantum Computing for Theoretical Nuclear Physics'' which was an outcome of the workshop {\it Quantum Computing for Nuclear Physics} held at the Institute for Nuclear Theory, University of Washington, during 14--15 November 2017~\cite{INT:QC}. The outline of the remainder of this document is as follows: Sec.~\ref{sec:qc} gives an overview of quantum computing, simulations, and qubit technologies; Sec.~\ref{sec:qs} discusses quantum sensors and other opportunities; and conclusions are provided in Sec.~\ref{sec:conclusions}

\section{Quantum Computing \label{sec:qc}}
\subsection{Background}
Quantum computing is a new paradigm for computing where attributes of quantum mechanics are leveraged to make possible new {\it quantum algorithms} that cannot be implemented efficiently on a classical Turing machine.\footnote{A quantum computer with around 300 qubits would contain more possible states than there are atoms in the universe, it is therefore conceivable that it would not be possible to implement certain quantum algorithms on any classical computer.} The fundamental unit of information in a quantum computer is a quantum bit (qubit), which is physically implemented by any two level quantum system.\footnote{Multi-level quantum systems, called {\it qudits}, are also possible.} A qubit can take on any superposition of the form $\alpha\left|0\right> + \beta\left|1\right>$, in contrast to a classical bit which must be exactly 0 or 1. This feature alone is not sufficient to make a quantum computer useful, since it resembles an analog computer where the fundamental unit of information is an analog voltage. Analog computers lack the basic error correction protocols which are integral to digital logic. Quantum error correction is possible however, where redundant storage of quantum information is used to protect against specific forms of noise or decoherence~\cite{Nielsen:2010aa}. Quantum error correction makes the dream of {\it universal} fault tolerant quantum computing a possibility.

The key feature of quantum mechanics that creates a quantum computer is entanglement between a set of qubits. The profound impact of entanglement is illustrated following Ref.~\cite{Weinberg:2015aa}; the general state for a set of $n$ potentially entangled qubits reads $\Psi = \sum_{s_1 s_2 \ldots s_n}\,\alpha_{s_1 s_2 \ldots s_n}\,\psi_{s_1 s_2 \ldots s_n}$ where $\alpha_{s_1 s_2 \ldots s_n}$ are complex numbers, the sum is over qubit states, and $\Psi$ is subject to the usual normalization condition: $\sum_{s_1 s_2 \ldots s_n} \left|\alpha_{s_1 s_2 \ldots s_n}\right|^2 = 1$. A quantum computer with $n$ entangled qubits therefore contains, and can act upon, $2^n-1$ independent complex numbers. In contrast, the state of a classical memory containing $n$ bits can be represented by a string of $n$ zeros and ones. Therefore, for a given number of qubits a quantum computer can operate on an exponentially greater amount of information than a classical computer with the same number of bits, or conversely, a classical digital computer needs an exponentially larger amount of memory to simulate a quantum computer.

{\it A priori} it is unclear whether simply having entanglement between qubits can in practice result in an exponential speed improvement of a quantum computer over its classical counterpart. It also does not appear to reduce the complexity of NP-complete problems. Further, entanglement is a necessary but not sufficient condition to achieve an exponential speed improvement of a quantum computer over its classical counterpart, as seen for instance in the Gottesman-Knill theorem. However, there are several quantum algorithms that have been proven to provide either exponential or polynomial speed improvements over any known classical algorithm, famous examples include Shor's factoring algorithm~\cite{Shor:1994jg} and Grover's database search algorithm~\cite{Grover:1996rk}.

Beside the difficulty of maintaining entanglement between a large number of qubits over an extended period of time, quantum computers also have intrinsic limitations. For example, while a quantum computer can store an exponential amount more information compared to an analogous classical computer, this information is in general not available to be read out from the quantum memory. A likely computing model for the foreseeable future is therefore having a classical CPU based computer as the computing hub, which then executes certain tasks on auxiliary quantum computers. A similar scenario already exists between CPUs and GPUs.

\subsection{Types of Quantum Computers}
There exist a number of different types of physical realizations of a quantum computer. The ultimate goal is to build a {\it universal} quantum computer, which for the purposes of this document can be considered a device that can execute any quantum algorithm. This can be defined as a quantum computer that can execute a complete set of universal quantum gates an arbitary number of times, where a set of gates is considered universal if any function in a given computational model can be computed to arbitary precision by a {\it circuit} that uses only these gates.

A universal gate set can usually be subdivided into gates that operate on a single qubit and those that operate on two or more qubits. The former class is used for the generation of superpositions, while the latter class is necessary for the generation of entanglement. The single qubit gates are usually $\sigma_x(\theta)$ and $\sigma_z(\theta)$ rotations about the $x$ and $z$ axes of the Bloch sphere. Because $\theta$ can take on any value, this class of gates is technically infinite, although it is usually considered sufficient to demonstrate a small number of such rotations. The rotation $\sigma_x(\theta)$ corresponds to a rotation between qubit levels, e.g. $\left|0\right> \to \cos(\theta)\left|0\right> + \sin(\theta)\left|1\right>$. Then $\sigma_z(\theta)$ rotations corresponds to a phase shift, e.g. $(\left|0\right> + \left|1\right>)/\sqrt{2} \to [\left|0\right> + \exp(i\theta)\left|1\right>]/\sqrt{2}$. 

There is a growing number of known multiple qubit gates that constitute full sets, where an important example is the controlled-not or CNOT gate. In a CNOT gate, the second qubit has a NOT operation, or $\sigma_x(\pi)$, performed on it if the first qubit is 1. Nothing is done if the first qubit is 0. A CNOT gate together with the two aforementioned single qubit rotations constitute a universal set of quantum gates~\cite{Nielsen:2010aa}. However, quantum error correction for qubit rotations by an arbitrary angle $\theta$ remains a significant challenge, and therefore only discrete rotations usually appear as part of a practical universal gate set.

Another important distinction is between {\it digital} and {\it analog} quantum computers. Digital quantum computers work by executing a discrete set of quantum operations (gates), which need not be a universal set of gates. Analog quantum computers or analog quantum simulators~\cite{Lloyd:1996aa,Buluta:2009aa} are purpose built quantum devices that are designed to simulate a specific class of quantum systems.  Analog quantum computers are generally not universal, however in the NISQ era, they may be the most efficient way to model aspects of quantum systems of interest -- such as lattice gauge theories~\cite{Zohar:2015hwa} -- in part because they do not rely on quantum error correction.

\subsection{Quantum Simulators}
A major motivation for quantum computing has always been the simulation of strongly coupled many-body quantum systems, with the reasoning that a quantum computer would already have built in the essential elements that make these calculations difficult on any classical computer.  A quantum simulator is a collection of qubits (or other multi-level quantum system) used to emulate the effects of a Hamiltonian that is precisely controlled by the experimenter.  In order for a quantum simulator to be more useful than the physical system it seeks to simulate, it must have some degree of control not available in the original system.  For example, one could possess the ability to prepare a system in a particular state, tune the Hamiltonian that governs dynamics, or measure the quantum state of the system.  The ultimate in controllability would be a full universal quantum computer, but these systems are limited in fidelity and qubit number. So we might ask if it is possible to achieve the desired simulation with less sensitivity to noise, even if that comes at the expense of operational control.  For example, the interactions between qubits could simply be tuned to have the same properties as the system under study, and allowed to evolve naturally.  In this case, the effect of decoherence will just appear as noise in the final result, and in fact may be useful for studying decoherence in the physical system being simulated.  This is closer to what is meant by an analog quantum simulator.

Analog quantum simulators are typically easier to implement in the lab than their digital counterparts: using the trapped ion platform as an example, the physical system can be engineered to behave like interacting quantum spins, with long-range interactions due to the underlying Coulomb interactions \cite{Zhang:2017aa}. For quantum computing experiments, great effort is devoted to tailoring them (e.g. through temporal pulse shaping) into arbitrary two-qubit gate operations \cite{Debnath2016}.  However, in an analog setting, the quantum simulation will exhibit the long-range behavior naturally \cite{Zhang:2017aa}, and hence could potentially be engineered to yield nuclear potentials with intrinsic long-range components, such as pion-exchange mediated forces between nucleons in effective field theories. Another example is the quantum simulation of the Schwinger model \cite{Martinez:2016yna}, which leveraged the naturally implemented long-range gates to produce the gauge degree of freedom within an ion chain. The primary advantage of a quantum simulation for nuclear many-body or gauge theory problems, therefore, is in reducing the qubit requirement of a fully digital computation, avoiding the costly discretization of some of the dynamical degrees of freedom by mapping them onto continuous interactions present in the simulator.

The simulations of an analog quantum simulator necessarily inherit some features of the physical qubit.  This can be advantageous - for instance, the N-fold spin symmetry (${\rm N}>1$) of cold atom systems is an essential feature used in proposals to simulate SU(N) and U(N) non-abelian gauge theories.  In other ways it can be disadvantageous, since the physical qubit must be chosen to match the system we wish to simulate.  Some examples of relevant systems include neutral atoms \cite{Gross2017}, trapped ions \cite{Blatt2012}, and artificial atoms such as superconducting qubits \cite{Houck2012}.  A universal quantum computer on the other hand will be much more flexible as the computation will be agnostic of the underlying implementation. The practical issue is whether the ``gate compiling'' can be done efficiently, so as to render a circuit that could run on near-term quantum machines, before the computation is spoiled by decoherence time and systematic errors.  From this perspective, analog systems could do better in the near to intermediate term due to the simultaneous nature of the interactions as well as larger system sizes.  It is possible that gate compiling challenges could be addressed with further investment in algorithms development.  Therefore, the co-existence of the two types of experimental approaches is crucial, and the development of analog-digital mixed approaches would assist in solving interesting problems in the NISQ era \cite{Preskill:2018aa}.

\subsection{Qubit Technologies}
The road to a full, universal, digital quantum computer that is consistently faster than its classical counterpart is probably decades long.  Moreover, we should not forget that quantum computers scale better than classical algorithms only for {\it specific} problems, so it may be that even a fully mature technology will not improve performance for most computational problems.  For these reasons, we should strategically seek out problems for which quantum algorithms and technologies have a strong advantage, so as to have the largest impact in the near to intermediate term, with an eye towards long-term success.  As we have seen, nuclear physics holds several such problems that, if solved, can have a real impact on our understanding of nature.

No one today is able to foresee what type of qubit will be most important to a future quantum computer, which may actually rely on a combination of qubit types, or indeed the key technologies may not yet even be invented. Qualities such as gate fidelity and coherence time need to be improved
for all existing qubit technologies, but it is not clear where the breakthroughs may occur.  Or for instance, hybrid qubits combine the strengths of different technologies to make a system capable of overcoming their individual weaknesses.  We should not therefore restrict research to any one qubit technology, because future progress on quantum computing could be dramatically stymied.  Moreover, as seen in the workshop, a broad-based approach to QIS has enabled many new technologies that have positive feedback to fundamental research in the form of quantum sensors.  Thus, it would be a disservice to the field to prematurely select a favored direction.  However, it may be appropriate to form a panel responsible for setting time lines or benchmarks to gauge progress towards problems of interest to DOE.
 
\subsubsection{Trapped Ions}
Trapped ions are an exemplar case for the synergy between academia, national labs, and industry in developing quantum technologies.  In fact, national labs have played a central role in trapped ion quantum computing from the beginning; the first demonstration of any 2-qubit quantum gate was performed at NIST-Boulder, and national labs such as Sandia continue to play a key role in developing the trap technology that is crucial to the huge progress in this field, as we heard in the presentation of Daniel Stick.  Further progress in this field will demand a level of engineering support often not available in a University setting, which is where national laboratories can serve as a bridge, assisting in the perfection of the technologies that will ultimately be needed as ion trap quantum computing makes the transition into industrial production and application.  We heard about the state of this exciting area from Chris Monroe's presentation, where he showed us the power of collaborative efforts between academia, national labs, and startup companies.  Trapped ions are also a good example of qubit technologies that can be adapted to basic physics research, as we heard from John Bollinger's presentation on how entangled ensembles of ions can be used as an extremely sensitive force and electric field sensor, with implications for tests of fundamental physics.  Finally, as discussed by Christine Muschik and Chris Monroe, trapped ions have also made excellent progress recently in the simulation of quantum field theories, such as 1+1 quantum electrodynamics.

\subsubsection{Neutral Atoms}
Ultracold neutral atom systems are tremendously rich in their quantum interactions, which has been exploited over years of research at many academic institutions to produce a variety of synthetic, designer Hamiltonians using such phenomena as Rydberg blockade, exchange interactions, and Feshbach resonances.  That richness makes neutral atom qubits ideally suited to analog quantum simulation, in addition to digital quantum computing. This is because once the synthetic Hamiltonian is implemented on the cold atom system, its ground state can be found, or a state can be prepared and evolved. One advantage is that we can naturally incorporate fermionic and bosonic degrees of freedom into this state, corresponding to fermionic and bosonic atomic species included in the lattice, as exhibited in the talk by Erez Zohar and also in Refs.~\cite{Zohar:2015hwa,Rico:2018pas}. Thus, by combining various controllable interactions, it is possible to build simulators to emulate unique quantum systems, which may someday include quantum chromodynamics.  Indeed, many of the extant proposals for simulating new gauge theories utilize this approach.  As has happened for trapped ions, there is an opportunity for DOE to advance neutral atom technology by fostering partnerships that bring together technological expertise from diverse fields.  In his presentation, Mark Saffman discussed one specific example, suggesting a partnership bringing spatial light modulation (SLM) technology expertise at ANL to bear on the problem of single qubit addressability.  The richness of cold atom interactions includes the possibility to form cold molecules, which is a young and exciting area of research.  Cold molecules can be used for enhanced searches for electric dipole moments (discussed in the quantum sensors section), or, as discussed by Bryce Gadway, used to enable new interactions for quantum simulators or computers.

\subsubsection{Superconducting Flux Qubits}
Qubits based on superconductors have made dramatic progress in recent years, and today they are probably the most important technology in the commercial space.  The ``cloud computing'' machines of IBM, Rigetti, and others are all based on this technology, and thus so is the quantum calculation of the deuteron mass presented by David Dean.  In comparison with trapped ion or neutral atom systems, superconducting flux qubits have extremely fast gate times, but comparatively short coherence times that typically limit their operational fidelity.  A large improvement in their coherence times would be transformational to quantum computing, but this will require difficult development in materials, microfabrication techniques, or superconducting technologies that may be beyond the capabilities of the university system.  These are areas of strength for national laboratories and DOE, so this presents a natural opportunity for collaboration that could have huge impacts on quantum computing, as discussed by Alex Romanenko and David Schuster in their presentations.

\subsubsection{Quantum Defects}
Quantum defects such as nitrogen vacancies have found many roles in quantum information science, such as single photon production for quantum cryptography, solid-state qubits, and tests of the foundations of quantum mechanics.  They are used for electric and magnetic field measurement, as well as microscopic thermometers.  These myriad applications suggests utility as quantum sensors, and indeed they have proposed to enhance precision measurements ranging from WIMP searches \cite{Rajendran2017} to chemical identification of picoliter volumes, such as proteins in a single living cell \cite{Glenn2018}.  Again, these devices are limited in coherence time, which is a problem best addressed as a materials purity concern.  One way nuclear physics could contribute to this area of research is in providing highly isotopically enriched samples, as outlined in the Isotopes section.

\section{Quantum Sensors \& Other Opportunities\label{sec:qs}}
A qubit, being simply an isolated and controllable two-level quantum system, is not only the basic unit of quantum computing, but really the most basic of quantum systems.  Thus, a variety of measurement techniques, including vapor cell magnetometry and nuclear magnetic resonance have qubits at their core. However, a true qubit is generally tuned to be highly insensitive to external interactions, whereas quantum sensors are instead tuned to be highly sensitive to specific fields or interactions.  This suggests that techniques originally developed to control qubits might bear fruit in measurement science.  One remarkable result is squeezing, which is a special case of Heisenberg-limited metrology, where quantum correlations between particles are used to reduce the shot noise limit \cite{Degen:2017aa}. In this case, sensitivity can improve linearly in the number of particles, rather than $\sqrt{N}$, and has been shown in specific cases to improve experimental sensitivities by 1 or 2 orders of magnitude.  Experiments that utilize these types of quantum sensors tend to be dominated by the demands of the sensor itself, to the point where often the quantum sensor {\it is} the experiment, as is the case for electric dipole moment measurements.  

Investment in QIS has repeatedly yielded residual benefits across basic science.  Frequently, we find technologies developed for QIS which may not satisfy the narrow definition of quantum sensors, but still have application elsewhere in the form of sensitive detectors, such as superconducting nanowires, nitrogen vacancy centers, or Rydberg atoms.  Through collaborative efforts to improve these technologies, QIS and NP can benefit from new investments in basic science.  Some opportunities are outlined below.

\subsection{Electric Dipole Moments}
Searches for electric dipole moments (EDMs) are perhaps the cleanest example of a quantum sensor applied to fundamental symmetries research.  We know that the observed baryon asymmetry of the universe demands that there be undiscovered sources of CP-violation hidden somewhere in nature, since the currently known sources of CP-violation are insufficient to create the abundance of normal matter in the universe.  EDMs are sensitive to CP-violation, and so non-zero EDMs are a generic feature of physics theories that extend the Standard Model.  Indeed, nature's bizarre preference for EDMs that are almost exactly zero has placed some of the strongest constraints on new physics theories ever since Ramsey performed the first neutron EDM experiment in the 1950s.  Today, a collection of EDM experiments together constrain the complicated phase space of CP-violating theories beyond the Standard Model, including the electron and  neutron, and EDMs from the nuclei of mercury, xenon, radium, and thallium fluoride.  These experiments generally work by measuring the Larmor precession frequency of some atom or nucleon in the presence of an electric field.  In this case, the electric field will cause a shift to the precession frequency if and only if the subject possesses an EDM.  Because it involves the conversion of a signal into a frequency, this technique enables experiments of exquisite precision and sensitivity.  That same aspect make these experiments strong candidates to benefit from quantum sensing techniques, such as spin squeezing and electron shelving, which can dramatically improve the detection of phase \cite{Hosten2016,Dietrich2010}.  The demonstrated gains with spin squeezing in recent years are impressive, giving signal to noise improvements of one to two orders of magnitude over the standard quantum limit \cite{Hosten2016}, with a commensurate improvement to EDM sensitivity to be expected.  Few experiments today can take immediate advantage of such improvements, however as the next generation of cold-atom and cold-molecule EDM experiments come on line, they will have the ability to take advantage of these techniques, vastly improving our understanding of CP-violation within the nucleus.  Another exciting direction with connections to QIS is the search for EDMs with cold molecules.  The effective electric field within a molecule is roughly 1000 times stronger than what can be achieved in the lab, with a similar improvement in the sensitivity reach of molecule-based EDM experiments.  Such improvements have already been realized for electron EDM searches, but are still in their infancy for nuclear-based experiments.

An alternative explanation for the CP problem is the axion.  Axions are a highly parsimonious theory, and could simultaneously explain the lack of CP-violation in the nucleus as well as the observed abundance of dark matter throughout the universe.  In this theory, the physical parameter describing CP-violation underwent spontaneous symmetry breaking in the early universe, thereby destroying the evidence of CP-violation while leaving behind a pseudo-Goldstone boson called the axion.  This new massive particle would then permeate the universe while interacting very weakly with ordinary matter, thus becoming the source of the dark matter we now observe.  The axion could cause a non-zero but oscillating EDM in nuclei if several requirements are satisfied: it must exist, it must be the constituent particle of dark matter, it must be ultralight, and the Earth must reside in a sizable patch of dark matter \cite{Stadnik2014}.  Since the average EDM is still zero, all previous experiments would have been ignorant of the non-zero EDM.  A new generation of experiments are now being designed that will be sensitive to this interesting physical model, such as CASPER and ARIADNE (presented by Andrew Geraci).  In fact limits have already been placed on this kind of dark matter using neutron EDM.  These experiments explore a very different region of phase space from more traditional WIMP searches, since the dark matter particles considered here have masses less than an electron volt, which would be completely undetectable in a nuclear recoil experiment due to the low momentum transfer.

\subsection{Superconducting Technologies}
The Department of Energy has a long history of supporting development in superconducting devices, which have taken a central stage in quantum computing and now present a strong opportunity for cross-pollination between NP and QIS.  Transition edge sensors (TESs) and superconducting nanowire single-photon detectors (SNSPDs) are two such technologies, presented by Aaron Miller and Clarence Chang.  These devices are made from thin films of superconducting material and exploit the fact that the superconducting phase transition is very narrow in order to detect single photons that have energies as high as MeV or as low as the microwave regime.  Because they entail a phase transition and the interaction is thermal, these devices have extremely high quantum efficiencies, 50\% for TESs and over 90\% for SNSPDs, and SNSPDs are also extremely fast, with timing resolutions of order 10 picoseconds.  TESs have even been shown to have excellent energy resolution, a rarity for detectors of photons with this energy.  These instruments have previously been used for quantum cryptography, where the high detection efficiencies are critical, but are now being leveraged by researchers at Argonne for use in fundamental symmetry studies and cosmology.  For example, TES's are used at the South Pole telescope to measure the cosmic microwave background.  They are also being built with energy resolution sufficient for use in the CUPID neutrinoless-double beta decay experiment, where they will help in the identification of decay events.  Superconducting detectors could also have promising applications in detectors for colliders, possibly addressing challenges in timing resolution, as outlined in Jose Repond's presentation.  Several technological challenges must first be overcome.  However, progress is being made to produce versions that are radiologically hard and resistant to high magnetic fields, which could open the door to possible SNSPD-based calorimeters suitable for detectors at locations such as the future electron ion collider.

Due to their structure, superconducting flux qubits tend to have coherence times of order 100 $\mu$s.  This is a major disadvantage of superconducting qubits when compared to atom or ion trap-based quantum computers which routinely have coherence times measured in seconds.  However, the DOE has world-class experience in building superconducting cavities with coherence times of many seconds.  If these technologies could be successfully interfaced, the combination could be transformative for superconducting qubit technology and quantum computing as a whole.  This is precisely the objective behind the University of Chicago-Fermi Lab qubit collaboration, discussed by David Schuster and Alex Romanenko, which seeks to create a quantum memory for long-term storage of qubits during computation, using transmons.  One challenge they seek to address is to produce superconducting cavities that retain their high quality factor when they contain only a few photons.  Feedback to basic physics research is very strong in this case, since similar cavity technology is used to search for Primakov decay of axion-like dark matter.  If an axion-like particle passes into a high-Q superconducting cavity that is resonant with the axion mass, a decay mode will be hugely enhanced, where the axion decays into two photons which can then be detected.  A number of experiments have been designed to exploit this technique, such as the ADMX and HAYSTAC collaborations, which complement the Oscillating-EDM method discussed earlier for detection of low mass dark matter.  National labs have considerable expertise in these areas, including design and manufacturing, that could prove beneficial to progress in quantum computing in industry as well as fundamental physics research.

National labs have considerable expertise in these areas, including design and manufacturing, that can accelerate development of these techniques.  Indeed, other quantum critical transitions, besides the superconducting one, may provide a general avenue for designing new quantum-limited detectors suitable for nuclear physics applications~\cite{Yang:2018dyv}.  Such detectors could prove beneficial to progress in quantum computing in industry as well as fundamental physics research. 

\subsection{Field Detection}
The oldest examples of quantum sensors, such as vapor cell magnetometers, are sensitive to magnetic or electric fields.  Some more recent examples include trapped ions (John Bollinger), Rydberg atoms (Georg Raithel), and quantum defects, such as nitrogen vacancy centers.  As Georg Raithel showed us, Rydberg atoms are being developed as a source of absolute calibration for RF field intensities, which has potential applications to measurements for DOE facilities.   Precision measurements on Rydberg atoms can also be used to improve constraints on the Rydberg constant, with direct relevance to the proton radius puzzle. Andrew Geraci presented on optically levitated nanospheres, which when cooled to their quantum ground state, have fantastic potential for tests of modified gravity at short distances.  John Bollinger showed us how highly entangled systems of trapped ions can be used not only for quantum simulation, but also how squeezing can perform extraordinarily precise measurements of force, with implications for searches of ultralight dark matter.  Measurement of forces and fields is a basic operation of precision measurement and tests of fundamental symmetries, and some of the most exciting developments today take advantage of QIS techniques.

\subsection{Isotopes}
There are several applications for isotopes in quantum information.  One is closely related to elimination of magnetic noise in so-called quantum defects, such as Si:P or nitrogen vacancy (NV) centers.  Quantum defects are optically active spin sites embedded in some substrate; carbon diamond for NV centers and silicon for phosphorus.  These devices have many applications in QIS, such as single photon generation, magnetic and electric field detection, or in quantum computing as qubits. Like superconducting qubits, these devices generally have extremely limited coherence times.  For these examples, however, it is known that the coherence time is limited by the presence of nearby isotopic impurities that possess a magnetic dipole, carbon-13 or silicon-29 to be specific.  By using a highly enriched substrate, the coherence times of these systems can be enhanced by orders of magnitude, a key requirement for several applications  \cite{Balasubramanian2009,Muhonen2014}.  It should be noted that in the past magnetic noise from these impurities have been controlled using techniques very familiar to nuclear physicists, including hyperpolarization and dynamic nuclear polarization \cite{Foletti2009}.  There are some substrates where this will remain true because no spin-0 nucleus exists for the constituant atoms, such as for GaAs substrates.  Isotopic impurities may be responsible for noise in other quantum systems as well - its been suggested that noise in superconducting flux qubits is caused by environmental radioactivity, which suggests qubit coherence time could be improved using the same approaches developed by nuclear physics for precision measurement, such as neutrinoless double beta decay.  Identification of needs and capabilities for isotopically pure samples may call for a roundtable discussion to coordinate efforts between QIS and NP experts.

There are some examples in ion trap quantum computing where a specific isotope is desirable due to its nuclear spin.  For example, there is only one naturally occurring isotope of calcium with nuclear spin, Ca-43, which has an abundance of only .135\%.  Thus, qubits based on calcium generally do not take advantage of long-lived nuclear coherence, because Ca-43's scarcity makes it difficult to trap.  A robust source of isotopically enriched Ca-43 would enable new qubits that are rarely used today.  Barium-133 is another interesting opportunity.  Spin 1/2 nuclei are especially desirable in ion trap QC due to the simplicity of state preparation and readout, while barium ions are desirable due to the long wavelength of its optical transitions, which facilitates long distance communication between ions.  Unfortunately, the stable isotopes of barium have either spin 0 or spin 3/2.  The only exception is the radioisotope barium-133 (halflife 10.551 years, spin 1/2), which has recently been demonstrated for use as a qubit \cite{Hucul2017}, and has the potential to overcome several long-standing limitations.  Universities and national labs can work closely in the creation, preparation and handling of such isotope samples.

\subsection{Ghost Imaging}
Ghost imaging is a technique that exploits particle correlations to produce an image of an object without having spatial resolution on the particles that scatter from the object itself \cite{Moreau2017}. This in principle can enable researchers to image an object using an entangled particle of different wavelength or even type than what the object is exposed to.  For instance, although an object might only be opaque in the infrared, it can be imaged with entangled visible photons, where cameras are much cheaper. This has been demonstrated with photons for many years, but more recently it and related techniques have been employed to exciting effect with electrons, having keV or MeV energies. The `quantum electron microscope' is a variant of this technique used to perform electron microscopy without damaging the surface imaged \cite{Kruit2016}. More recently, MeV energy electrons were used to image an aperture without any spatial resolution on the high energy electron \cite{Li2018}. What if this technology could be extended in to the GeV energy range relevant to medium and high energy physics? Ghost imaging could enable entirely new detector or source schemes, where the relevant degrees of freedom of the electron wave function are mapped onto initial or entangled final states that are more easily accessed.

\subsection{Data Analysis}
The ultimate purpose of quantum computing is to build devices that are capable of executing algorithms faster than any classical algorithm achieving the same task.  These algorithms are necessarily quantum in nature, and cannot always be understood using the traditional paradigms of classical computing, which inherently lack the basic operations that enable the quantum speedup.  The types of problems that are known to have an efficient quantum analog are relatively small in number, although some of them potentially have very general use, such as the quantum Fourier transform, Grover's algorithm for database search, and matrix inversion \cite{Nielsen:2010aa,Harrow2009}.  If these known algorithms could be used in concert with modern computers, like a quantum co-processor, it could present an opportunity for an early application of quantum computing.  With the right choice of algorithm, nuclear physics could benefit substantially in the form of improved data analysis capabilities.  Clearly, technical challenges remain, not only in terms of demonstrating quantum supremacy for a specific problem, but also in terms of transferring data onto the co-processor efficiently, and managing the small number of noisy qubits which are likely to be available in the near to intermediate time frame.  Even if a processor with quantum supremacy is demonstrated, this input/output problem must be solved in order to not obviate the advantage the co-processor enables in the first place.  That sort of challenge is one where collaborations of the kind described here could excel, with engineering and computing expertise provided in conjunction with universities and private industry capabilities.  Due to the enormous significance of such technology, including to nuclear physics, a compelling data analysis application could mobilize many researchers within DOE to work on this topic.

\section{Conclusion\label{sec:conclusions}}
The road to universal quantum computing is just beginning but is likely paved with numerous opportunities which will further basic science and nuclear physics. These opportunities include potentially transformative new approaches to the calculation and simulation of nuclear systems, and new detectors for nuclear physics experiments. As we have discussed, QIS can address several challenges faced by nuclear physics, and QIS in turn will benefit from development through real-world problems. Incubation of these technologies for applications in physics will have broad consequences not only for our understanding of nature but also for computation generally. To strategically identify relevant problems and technologies a new community working at the intersection of nuclear physics and quantum information should be fostered, who can carefully identify and exploit these opportunities.  The myriad and significant technical challenges will demand the combined capabilities of academia, national laboratories, and industry, but their solution can bear rewarding fruit for both QIS and nuclear physics.  Herein, we have outlined a starting point for how such a collaboration may begin and some of the promising directions it could pursue.

\section*{Acknowledgements}
ICC and MRD would like to thank all of the organizers and participants of the workshop {\it Intersections between Nuclear Physics and Quantum Information} which lead to this whitepaper. This work was supported by the U.S. Department of Energy, Office of Science, Office of Nuclear Physics, contract no. DEAC02-06CH11357.

\addcontentsline{toc}{section}{References}
\printbibliography 
\appendix
\renewcommand{\thesection}{\Alph{section}.} 
\clearpage
\section{NPQI Workshop Participants and Agenda}
The website for the workshop {\it Intersections between Nuclear Physics and Quantum Information} can be found at \url{http://www.phy.anl.gov/npqi2018/} which, {\it inter alia}, contains a record of the presentations. The following gives the workshop participants and agenda: \\[0.9em]
\textbf{Yuri Alexeev}         Argonne National Laboratory,
\textbf{Whitney Armstrong}    Argonne National Laboratory,
\textbf{John Arrington}       Argonne National Laboratory,
\textbf{Birger Back}          Argonne National Laboratory,
\textbf{Alexei Bazavov}       Michigan State University,
\textbf{Silas Beane}          University of Washington,
\textbf{Kyle Bednar}          Kent State University,
\textbf{Leo Bellantoni}       Fermilab,
\textbf{Paul Benioff}         Argonne National Laboratory,
\textbf{Alexey Bezryadin}     University of Illinois,
\textbf{Michael Bishof}       Argonne National Laboratory,
\textbf{David Blyth}          Argonne National Laboratory,
\textbf{John Bollinger}       NIST,
\textbf{Richard Brower}       Boston University,
\textbf{Paul Bruillard}       PNNL,
\textbf{Mary Burkey}          University of Chicago,
\textbf{Michael Carpenter}    Argonne National Laboratory,
\textbf{Clarence Chang}       Argonne National Laboratory,
\textbf{Yong Chen}            Purdue University,
\textbf{Ian Clo\"et}          Argonne National Laboratory,
\textbf{Zachary Conway}       Argonne National Laboratory,
\textbf{David Dean}           Oak Ridge National Laboratory,
\textbf{Marcel Demarteau}     Argonne National Laboratory,
\textbf{Matthew Dietrich}     Argonne National Laboratory,
\textbf{Patrick Dreher}       NC State University,
\textbf{Eugene Dumitrescu}    Oak Ridge National Laboratory,
\textbf{Estia Eichten}        Fermilab,
\textbf{Eden Figueroa}        Stony Brook University,
\textbf{Mael Flament}         Stony Brook University,
\textbf{Stefan Floerchinger}  Heidelberg University,
\textbf{Aaron Fluitt}         Argonne National Laboratory,
\textbf{Nikolaos Fotiadis}    Los Alamos National Laboratory,
\textbf{Adam Freese}          Argonne National Laboratory,
\textbf{Bryce Gadway}         University of Illinois U-C,
\textbf{Dave Gaskell}         Jefferson Lab,
\textbf{Donald Geesaman}      Argonne National Laboratory,
\textbf{Andrew Geraci}        Northwestern University,
\textbf{Walter Giele}         Fermilab,
\textbf{Alexey Gorshkov}      Joint Quantum Institute,
\textbf{Anna Grassellino}     Fermilab,
\textbf{Stephen Gray}         Argonne National Laboratory,
\textbf{Nathan Guisinger}     Argonne National Laboratory,
\textbf{Peter Gysbers}        TRIUMF,
\textbf{Kawtar Hafidi}        Argonne National Laboratory,
\textbf{Mohammad Hattawy}     Argonne National Laboratory,
\textbf{Eric Herbert}         Lewis University,
\textbf{Joseph Heremans}      Argonne National Laboratory,
\textbf{Kiel Howe}            Fermilab,
\textbf{Ciaran Hughes}        Fermilab,
\textbf{Zubin Jacob}          Purdue University,
\textbf{Xiao-Yong Jin}        Argonne National Laboratory,
\textbf{Sereres Johnston}     Argonne National Laboratory,
\textbf{Bertus Jordaan}       Stony Brook University,
\textbf{David Kaplan}         University of Washington,
\textbf{Natalie Klco}         University of Washington,
\textbf{Andy Kowalski}        Jefferson Lab,
\textbf{Ushma Kriplani}       Argonne National Laboratory,
\textbf{T.-S. Harry Lee}      Argonne National Laboratory,
\textbf{Daniel Lopez}         Argonne National Laboratory,
\textbf{Ivar Martin}          Argonne National Laboratory,
\textbf{Jose Martinez}        Argonne and IIT,
\textbf{Edward May}           Argonne National Laboratory,
\textbf{Anna McCoy}           TRIUMF,
\textbf{Michael McGuigan}     Brookhaven National Laboratory,
\textbf{Robert McKeown}       Jefferson Lab,
\textbf{Yannick Meurice}      University of Iowa,
\textbf{Aaron Miller}         Quantum Opus LLC,
\textbf{Misun Min}            Argonne National Laboratory,
\textbf{Peter Mintun}         University of Chicago,
\textbf{Christopher Monroe}   University of Maryland,
\textbf{Peter Mueller}        Argonne National Laboratory,
\textbf{Christine Muschik}    Institute for Quantum Computing,
\textbf{Petr Navratil}        TRIUMF,
\textbf{Mike Norman}          Argonne National Laboratory,
\textbf{Blaine Norum}         University of Virginia,
\textbf{Thomas O'Donnell}     Virginia Tech Physics Department,
\textbf{James Osborn}         Argonne National Laboratory,
\textbf{Matthew Otten}        NanoScience and Technology,
\textbf{Peter Petreczky}      Brookhaven National Laboratory,
\textbf{Tomas Polakovic}     Argonne National Laboratory,
\textbf{Alan Poon}            Berkeley Lab,
\textbf{Raphael Pooser}       Oak Ridge National Laboratory,
\textbf{David Potterveld}     Argonne National Laboratory,
\textbf{John Preskill}        Caltech,
\textbf{Tenzin Rabga}         Argonne National Laboratory,
\textbf{Gulshan Rai}          U.S. Department of Energy,
\textbf{Georg Raithel}        University of Michigan,
\textbf{Thomas Reid}          Argonne National Laboratory,
\textbf{Jose Repond}          Argonne National Laboratory,
\textbf{Pedro Rivero}         Illinois Institute of Technology,
\textbf{Alessandro Roggero}   Los Alamos National Laboratory,
\textbf{Alexander Romanenko}  Fermilab,
\textbf{Mark Saffman}         University of Wisconsin-Madison,
\textbf{Daniel Santiago}      Argonne National Laboratory,
\textbf{Martin Savage}        Institute for Nuclear Theory,
\textbf{John Schiffer}        Argonne National Laboratory,
\textbf{David Schuster}       University of Chicago,
\textbf{James Simone}         Fermilab,
\textbf{Pete Slawniak}        Argonne National Laboratory,
\textbf{Paul Sorensen}        U.S. Department of Energy,
\textbf{Daniel Stick}         Sandia National Labs,
\textbf{Oleksii Strelchenko}  Fermilab,
\textbf{Brent VanDevender}    PNNL,
\textbf{Frank Verstraete}     University of Ghent,
\textbf{Matteo Vorabbi}       TRIUMF,
\textbf{Gensheng Wang}        Argonne National Laboratory,
\textbf{Robert Wiringa}       Argonne National Laboratory,
\textbf{Junqi Xie}            Argonne National Laboratory,
\textbf{Liang Yang}           University of Illinois U-C,
\textbf{Yongchao Yang}        Argonne National Laboratory,
\textbf{Boram Yoon}           Los Alamos National Laboratory,
\textbf{Linda Young}          Argonne National Laboratory,
\textbf{Jake Zappala}         Argonne National Laboratory,
\textbf{Jiehang Zhang}        University of Maryland,
\textbf{Zhaohui Zhang}        Northeastern University,
\textbf{Erez Zohar}           Max Planck Institute of Quantum Optics.

\begin{figure*}[ht!]
\centering\includegraphics[width=1.0\textwidth]{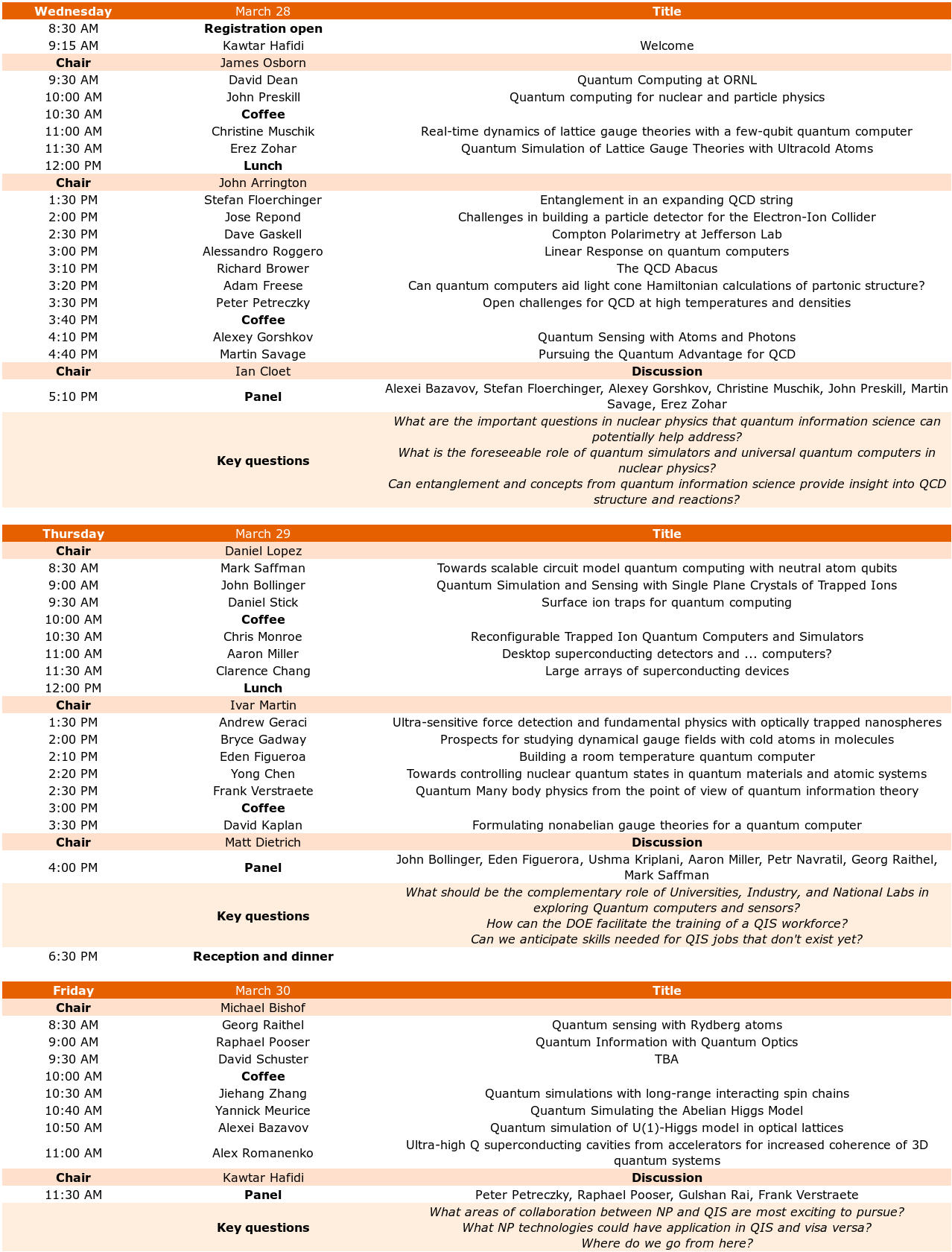}
\end{figure*}

\newpage


\section*{Supplementary material}

# A. NPQI WORKSHOP PARTICIPANTS AND AGENDA

| Wednesday | March 28 | Title |
|---|---|---|
| 8:30 AM | **Registration open** | |
| 9:15 AM | Kawtar Hafidi | Welcome |
| **Chair** | James Osborn | |
| 9:30 AM | David Dean | Quantum Computing at ORNL |
| 10:00 AM | John Preskill | Quantum computing for nuclear and particle physics |
| 10:30 AM | **Coffee** | |
| 11:00 AM | Christine Muschik | Real-time dynamics of lattice gauge theories with a few-qubit quantum computer |
| 11:30 AM | Erez Zohar | Quantum Simulation of Lattice Gauge Theories with Ultracold Atoms |
| 12:00 PM | **Lunch** | |
| **Chair** | John Arrington | |
| 1:30 PM | Stefan Floerchinger | Entanglement in an expanding QCD string |
| 2:00 PM | Jose Repond | Challenges in building a particle detector for the Electron-Ion Collider |
| 2:30 PM | Dave Gaskell | Compton Polarimetry at Jefferson Lab |
| 3:00 PM | Alessandro Roggero | Linear Response on quantum computers |
| 3:10 PM | Richard Brower | The QCD Abacus |
| 3:20 PM | Adam Freese | Can quantum computers aid light cone Hamiltonian calculations of partonic structure? |
| 3:30 PM | Peter Petreczky | Open challenges for QCD at high temperatures and densities |
| 3:40 PM | **Coffee** | |
| 4:10 PM | Alexey Gorshkov | Quantum Sensing with Atoms and Photons |
| 4:40 PM | Martin Savage | Pursuing the Quantum Advantage for QCD |
| **Chair** | Ian Cloet | **Discussion** |
| 5:10 PM | **Panel** | Alexei Bazavov, Stefan Floerchinger, Alexey Gorshkov, Christine Muschik, John Preskill, Martin Savage, Erez Zohar |
| | **Key questions** | *What are the important questions in nuclear physics that quantum information science can potentially help address?*<br>*What is the foreseeable role of quantum simulators and universal quantum computers in nuclear physics?*<br>*Can entanglement and concepts from quantum information science provide insight into QCD structure and reactions?* |

| Thursday | March 29 | Title |
|---|---|---|
| **Chair** | Daniel Lopez | |
| 8:30 AM | Mark Saffman | Towards scalable circuit model quantum computing with neutral atom qubits |
| 9:00 AM | John Bollinger | Quantum Simulation and Sensing with Single Plane Crystals of Trapped Ions |
| 9:30 AM | Daniel Stick | Surface ion traps for quantum computing |
| 10:00 AM | **Coffee** | |
| 10:30 AM | Chris Monroe | Reconfigurable Trapped Ion Quantum Computers and Simulators |
| 11:00 AM | Aaron Miller | Desktop superconducting detectors and … computers? |
| 11:30 AM | Clarence Chang | Large arrays of superconducting devices |
| 12:00 PM | **Lunch** | |
| **Chair** | Ivar Martin | |
| 1:30 PM | Andrew Geraci | Ultra-sensitive force detection and fundamental physics with optically trapped nanospheres |
| 2:00 PM | Bryce Gadway | Prospects for studying dynamical gauge fields with cold atoms in molecules |
| 2:10 PM | Eden Figueroa | Building a room temperature quantum computer |
| 2:20 PM | Yong Chen | Towards controlling nuclear quantum states in quantum materials and atomic systems |
| 2:30 PM | Frank Verstraete | Quantum Many body physics from the point of view of quantum information theory |
| 3:00 PM | **Coffee** | |
| 3:30 PM | David Kaplan | Formulating nonabelian gauge theories for a quantum computer |
| **Chair** | Matt Dietrich | **Discussion** |
| 4:00 PM | **Panel** | John Bollinger, Eden Figuerora, Ushma Kriplani, Aaron Miller, Petr Navratil, Georg Raithel, Mark Saffman |
| | **Key questions** | *What should be the complementary role of Universities, Industry, and National Labs in exploring Quantum computers and sensors?*<br>*How can the DOE facilitate the training of a QIS workforce?*<br>*Can we anticipate skills needed for QIS jobs that don't exist yet?* |
| 6:30 PM | **Reception and dinner** | |

| Friday | March 30 | Title |
|---|---|---|
| **Chair** | Michael Bishof | |
| 8:30 AM | Georg Raithel | Quantum sensing with Rydberg atoms |
| 9:00 AM | Raphael Pooser | Quantum Information with Quantum Optics |
| 9:30 AM | David Schuster | TBA |
| 10:00 AM | **Coffee** | |
| 10:30 AM | Jiehang Zhang | Quantum simulations with long-range interacting spin chains |
| 10:40 AM | Yannick Meurice | Quantum Simulating the Abelian Higgs Model |
| 10:50 AM | Alexei Bazavov | Quantum simulation of U(1)-Higgs model in optical lattices |
| 11:00 AM | Alex Romanenko | Ultra-high Q superconducting cavities from accelerators for increased coherence of 3D quantum systems |
| **Chair** | Kawtar Hafidi | **Discussion** |
| 11:30 AM | **Panel** | Peter Petreczky, Raphael Pooser, Gulshan Rai, Frank Verstraete |
| | **Key questions** | *What areas of collaboration between NP and QIS are most exciting to pursue?*<br>*What NP technologies could have application in QIS and visa versa?*<br>*Where do we go from here?* |

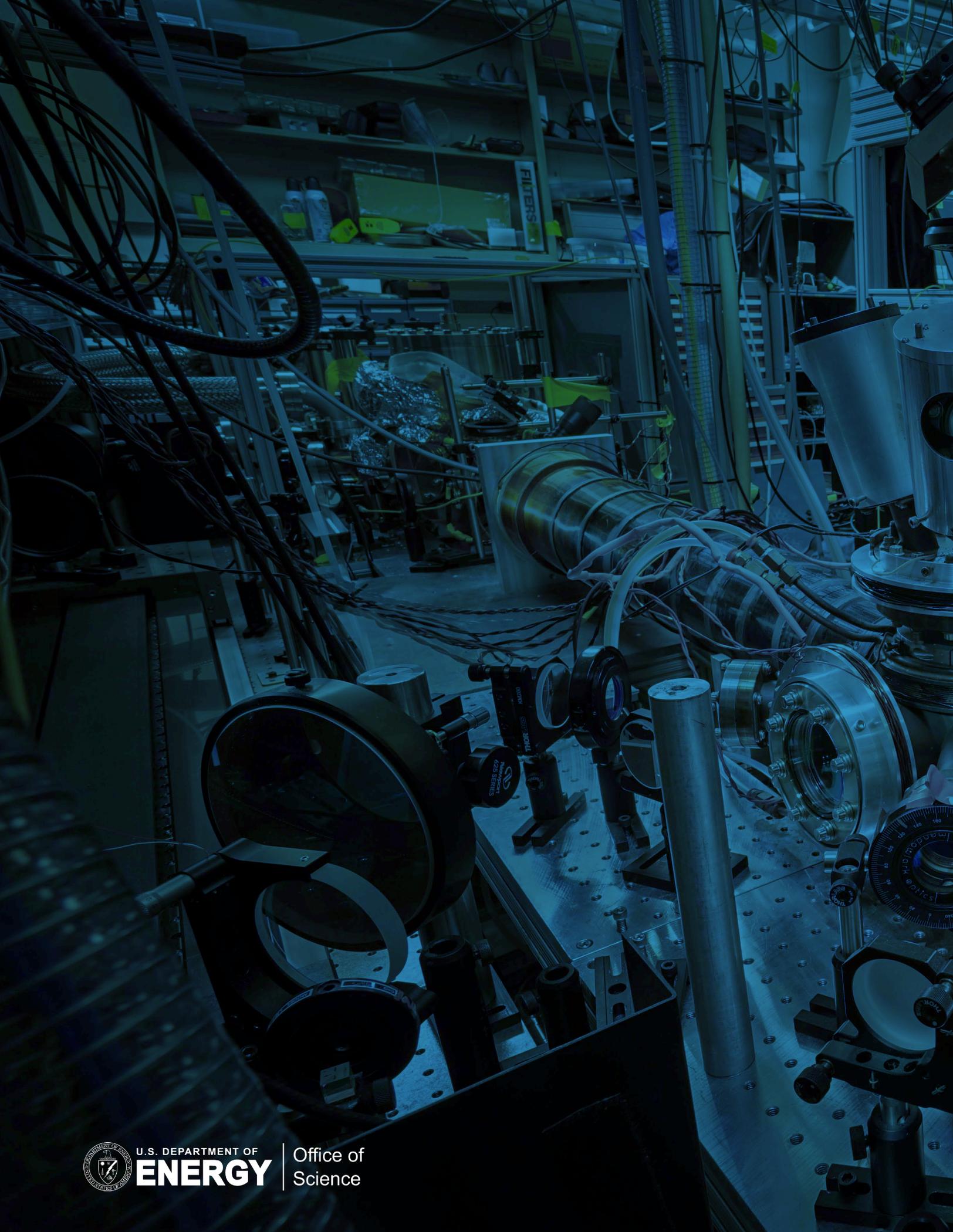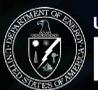


\end{document}